\documentstyle[preprint,aps,prb]{revtex}
\begin{document}

\draft

\title{
Extended Aharonov-Bohm period analysis of strongly correlated 
electron systems
}

\author{ Ryotaro {\sc Arita}$^1$, Koichi {\sc Kusakabe}$^2$, 
Kazuhiko {\sc Kuroki}$^1$, and Hideo {\sc Aoki}$^1$ } 
\address{$^1$Department of Physics, University
 of Tokyo, Hongo, Tokyo 113, Japan} 
\address{$^2$Institute for Solid
  State Physics, University of Tokyo, Roppongi, Tokyo 106, Japan}
\date{\today}

\maketitle
\begin{abstract}
The `extended Aharonov-Bohm (AB) period' recently
proposed by Kusakabe and Aoki [J. Phys. Soc. Jpn {\bf 65}, 2772
(1996)]
is extensively studied numerically for finite size systems 
of strongly correlated electrons.
While the extended AB period is the system length times the flux quantum 
for noninteracting systems, 
we have found the existence of the boundary across which 
the period is halved or another boundary into an even shorter 
period on the phase diagram for these models.   
If we compare this result with the phase diagram predicted from 
the Tomonaga-Luttinger theory, devised for low-energy physics, 
the halved period (or shorter periods) has a one-to-one 
correspondence to the existence of the 
pairing (phase separation or metal-insulator 
transition) in these models. 
We have also found for the $t$-$J$ model that the extended AB period does not
change across the integrable-nonintegrable boundary 
despite the totally different level statistics.
\end{abstract}

\medskip


\section{Introduction}

Aharonov-Bohm (AB) effect is an interesting probe for 
various phenomena in 
condensed-matter physics, since this effect explores 
`global' responses (over large changes in the gauge flux) 
rather than linear responses against an infinitesimal change.  
In this light, it should be especially interesting to look at 
strongly correlated systems such as the Hubbard model or the 
$t$-$J$ model from the AB effect.   
A natural interest is how the occurrence of 
superconductivity, if any, should be reflected in the AB effect.  
The response of the system to the AB flux 
has long been recognized as 
the `anomalous flux quantization' as 
a tool for detecting the Cooper pairing ever since 
Byers and Yang proposed the idea.\cite{byers} 
Let us pierce an AB flux, $\Phi$, through the torus
as shown in Figure.\ref{torus}.
In a normal state, the ground-state energy is a periodic
function of $\Phi$ with period
$\Phi_0 =hc/e$ with minima at
$\Phi/\Phi_0= 0,\pm 1, \pm 2, \cdots$.  
In a superconducting phase, new stable states emerge 
at $\Phi/\Phi_0=\pm \frac{1}{2}, \pm \frac{3}{2} \cdots$ 
due to the pairing, which makes the period 
$\Phi_0/2$.  
The method has been applied to several models\cite
{russian1,russian2,russian3,russian4,russian5,russian6,Ferretti}.

The response to an 
infinitesimal 
AB flux can also detect 
the Drude weight $D$\cite{ss} 
or the superfluid weight $D_s$.\cite{swz1,swz2} 
The Drude weight is a measure of the ratio of the density
of mobile charge carriers to their mass, while the 
superfluid weight measures the ratio of the superfluid density to mass. 
For a clean system, we expect 
$D=D_s=0$ for an insulator, $D$ is finite and
$D_s=0$ for a metal, and $D$ and $D_s$ are both finite 
for a superconductor. 

As a test for pairing, however, the conventional 
anomalous flux quantization 
can be ambiguous, 
since the method is, in some cases, 
not sensitive enough to determine phase boundaries, 
or because finite-size effects obscure the test. 
For example, Ferretti {\it et al.} shows 
that both repulsive and attractive Hubbard model behave as
a superconductors in the sense of Byers-Yang arguments at $T$=0.
\cite{Ferretti}  
We can see this in the left corner of Fig.\ref{kusakabe}.

In order to shed a new light, two of the present authors 
have recently proposed an entirely different 
method of detecting bound electron states
(see Fig.\ref{kusakabe})
.\cite{Kusakabe} 
The idea stemmed from 
Sutherland's study for 
one-dimensional (1D) 
Heisenberg magnets.\cite{Sutherland} 
He looks at what happens to a many-body state when we
twist the boundary condition as  

\[\phi(\cdots,x_{j}+N_{a},\cdots)=e^{i2\pi \Phi/\Phi_0}\phi(\cdots, 
x_{j},\cdots) \, , \]

\noindent for $N_a$-site lattice 
shifts (or `boosts') the total momentum of an $N$ particle state.  
For a noninteracting particles, the boost by $\Phi =\Phi_0$ 
shifts each one-particle momentum by $2\pi/N_a$, so that 
$\Phi=N_a \Phi_0$ will bring the set of $k$ points 
back to the original position.

For an interacting particles we can expect 
that $\Phi=N_a \Phi_0$ will also recover the 
original eigenstate 
when all of the $N$ particles move individually. 
By contrast, the response will be different 
if the particles form $N$-bound states, 
since bound $N$ particles should have a well-defined center-of-mass 
momentum, whose response to the AB flux will be $N$ times greater. 
A twist of $\Phi=N_a \Phi_0/N$ will then suffice to bring the energy back to 
the initial value. 

For the Heisenberg spin chain Sutherland\cite{Sutherland} 
has in fact confirmed this behavior 
using the Bethe-ansatz analysis. 
Kusakabe and Aoki\cite{Kusakabe} have 
then proposed that the idea for spin systems 
may be extended 
to electron systems, such as the Hubbard model, 
for detecting bound-electron states.  
In addition to the Bethe-ansatz analysis for electrons, 
the method is also numerically implemented 
in order to reveal the phase diagram of strongly correlated systems
in (pseudo-)1D and possibly in 2D systems, 
where the Bethe ansatz is
inapplicable but we can still keep track of the 
$\Phi$-dependence of the state over the 
`extended zone' $(0 \leq \Phi \leq N_a \Phi_0)$ well beyond 
the one period $(0 \leq \Phi \leq \Phi_0)$. 

Thus this technique is called `the extended AB period method'. 
It is conceptually interesting to keep track of the states along the 
extended zone, since this questions an even more global response of 
a many-body system than the anomalous flux quantization.

With this method a normal-to-superconducting transition is expected 
to be characterized 
as a halving of the extended period. 
For the 1D Hubbard model, which is an integrable system, 
this has been shown indeed to be the case.\cite{Kusakabe} 
The spectral flows have also been  analyzed in 
other integrable systems by Fukui and Kawakami 
in the context of the exclusion statistics, such as 
the Haldane-Shastry model
\cite{Fukui1} and the supersymmetric $t$-$J$ model 
with $1/r^2$ interaction.\cite{Fukui2}
On the other hand, the numerical implementation has been 
applied to non-integrable systems.\cite{Kusakabe,Kusakabe2} 

In the present paper, 
we investigate extensively the extended AB method 
by applying it to 
one-dimensional 
(1) $t$-$J$ model, 
(2) $t$-$J$ ladder model, 
(3) $t$-$J$-$J'$ model, and 
(4) the extended Hubbard model 
to systematically look at the behavior of the extended period.

We address ourselves two fundamental questions: 

(i) what is the difference 
between integrable and non-integrable systems 
in terms of the extended AB flow, 

(ii) can we actually detect normal-superconducting 
transition or phase separation 
in such nontrivial models as enumerated above,
where a quantum phase transition (at $T=0$) is believed 
to occur at a finite strength of the interaction. 

As for the first question, we can naively expect 
drastically different behaviors for the two 
classes of systems, since 
the extended AB period is dominated by 
whether the level crossings in the adiabatic flow line turn into 
anti-crossings, while 
the quantum adiabatic theorem dictates that the adiabatic 
evolution of a level 
is free from level crossings 
(except for accidental ones) when the system is nonintegrable.   
We can in fact characterize the avoided crossings in terms of 
the level statistics (distribution of the spacing of adjacent 
eigenenergies), which is thought to be universally 
Gaussian orthogonal (unitary or 
symplectic depending on the symmetry of the Hamiltonian) 
for a nonintegrable system while an integrable 
system has 
Poisson's distribution as in a noninteracting system.  
This is because, while levels usually have to repel each other 
except for accidental degeneracies in the non-integrable case, 
there exist enough symmetries to 
characterize a state that allows the levels to cross 
in the integrable case. 
This should also apply to many-body systems in integrable or 
nonintegrable cases.  
For instance, Di Stasio and Zotos\cite{Zotos} have obtained level statistics 
for spinless fermions in one dimension with nearest and
next-nearest-neighbor interaction.   
Thus it is intriguing whether the extended spectral flow analysis, 
which has been shown to be useful in integrable systems from the Bethe-Ansatz 
analysis, could be applied to non-integrable systems as well. 
We find here 
that the spectral flow is smooth in 
solvable cases as opposed to wiggly flows in nonintegrable cases, 
but that its period does not change across 
the integrable-nonintegrable boundary.

For the question (ii) we conclude here that 
the extended AB method can indeed detect a 
symptom of changes in characters of quantum states
even for non-integrable system. 
For the 1D $t$-$J$ model, the phase diagram 
has been obtained by Ogata {\it et al.}\cite{Ogata} by 
combining the Tomonaga-Luttinger (TL) theory\cite{Schulz,Kawakami,Frahm} 
and the Lanczos diagonalization 
of finite systems.
With the same method, Ogata {\it et al.}\cite{tJJ}
obtained the phase diagram of 1D $t$-$J$-$J'$ model, and
Sano and {\=O}no \cite{Sanoeh},
and independently Kuroki {\it et al} \cite{kurokieh},
obtained the phase diagram of extended Hubbard model.
For the $t$-$J$ ladder, several authors have obtained the phase diagram
using the TL theory assuming that the ladder system may be described 
by a 1D effective theory.\cite{Hayward96,Sano96,Tsunetsugu95}
The phase diagrams obtained from the extended AB results 
are qualitatively consistent with 
those predicted by TL theory.

The organization of the paper is as follows.
In section 2 we
describe the numerical implementation for keeping track of the 
extended flow.  
We shall then discuss the result for the 1D $t$-$J$ model in section 3, 
where we perform the Bethe-ansatz analysis to discuss the origin of 
the extended periodicity. 
Specifically, we discuss the behavior of the extended AB period
in the non-integrable region from a 
numerical analysis. 
Although the period does not change across the integrable-nonintegrable 
boundary, 
their difference does appear 
in line shapes of the spectral flow.
We discuss the results for the $t$-$J$ ladder (Section 4), 
1D $t$-$J$-$J'$ model (Section 5) and extended Hubbard model
(Section 6).

\section{Numerical Algorithm}
In general we want to treat non-integrable systems.  
For instance 
the $t$-$J$ model is non-integrable even in 1D except at 
special points, i.e., $J=0$ ($U=\infty$ Hubbard model) 
or $J=2t$ (the supersymmetric $t$-$J$ model). 
Thus we must numerically implement the method to obtain the 
extended AB period. 
Here we recapitulate the numerical algorithm proposed 
in Ref.12,
which consists of successive estimations of the energy and wave function. 
\begin{enumerate}
\item 
First solve $H(\Phi=0)\phi=E(\Phi=0)\phi$ by a conventional diagonalization 
technique, ({\it e.g.,} the Lanczos method). 
\item
Predict $\tilde{E}_{n=0}(\Phi+\Delta \Phi)$ by 
$\tilde{E}_{n=0} \equiv 
\langle \phi(\Phi)|H(\Phi+\Delta \Phi)|\phi(\Phi)\rangle$. 
\item
Determine $\phi_{n+1}(\Phi+\Delta \Phi)$ by the power method, 
\[
  \left[\frac{1}{H(\Phi+\Delta \Phi)-\tilde{E}_{n}}\right]^m \phi(\Phi) 
  \rightarrow \phi_{n+1}(\Phi+\Delta \Phi)
\]
\item 
Determine $\tilde{E}_{n+1}(\Phi+\Delta \Phi)$ by 
$H(\Phi +\Delta \Phi) \phi_{n+1}(\Phi +\Delta \Phi)= 
\tilde{E}_{n+1}(\Phi +\Delta \Phi) \phi_{n+1}(\Phi +\Delta \Phi)$. 
\item
If a convergence criterion, 
$|\tilde{E}_{n+1}(\Phi +\Delta \Phi) - \tilde{E}_{n}(\Phi +\Delta \Phi)| 
\le \varepsilon$ is met with typically $\varepsilon \sim 10^{-8}$, 
then increase $\Phi$ and go to 2. 
Otherwise, increase $n$ and go to 3. 
\end{enumerate}
Usually the energy converges faster than the wavefunction, 
so that the above algorithm works well. 
Typically, a few tens of $m$ and $n$ give a convergence. 

Around level crossings a care must be taken.  
There are two types of level crossings. 
One type occurs between levels that have 
different symmetries. 
In this case we can readily go across the crossing, since our method 
looks at the wavefunction, which is a representation of all of 
the symmetries of the system. 
The second type occurs at a critical point, at which 
a gap opens in the spectral flow. 
Thus we must question how a crossing turns 
into an anti-crossing from crossing at 
a critical point. 
We can systematically keep track of this 
by varying interaction parameters around the critical point. 

\section{1D $t$-$J$ model}

We first analyze the extended AB period of the 
1D $t$-$J$ ring, where the main interest is 
whether a change in the period detects a symptom 
in characters of quantum states.
The phase diagram of this model has been investigated by 
the Tomonaga-Luttinger theory combined with numerical 
results,\cite{Ogata} 
with which we can compare our result. 

The $t$-$J$ model belongs to a class of the strongly correlated 
model in which the electron repulsion is assumed to be infinite so that 
no double occupancies of electrons are allowed.  
The Hamiltonian is given by
\[
  H=-t \sum_{\langle i,j\rangle \sigma} (\tilde{c}^{\dagger}_{i \sigma} 
\tilde{c}_{j \sigma} 
  + {\rm H.c.}) + J \sum_{\langle i,j\rangle }
  ({\bf S}_{i}\cdot {\bf S}_{j}-\frac{1}{4} n_{i}n_{j}) \; , 
\]
where $\tilde{c}^{\dagger}_{i \sigma}$ creates at the $i$-th site
with spin $\sigma$ and 
$S_{i}$ is the spin operator defined by 
$S_{i}=\frac{1}{2} \tilde{c}_{i\sigma}^\dagger \vec{\sigma}_{\sigma \sigma'} 
\tilde{c}_{i\sigma'}$ with the Pauli matrix $\vec{\sigma}_{\sigma \sigma'}$. 
Hereafter we set $t=1$ and the phase diagram will be 
parameterized by $J$ and the density of electrons. 

Figure \ref{224}, \ref{226}, \ref{228}
shows the numerical result for 
the spectral flow 
of the 1D $t$-$J$ ring with various values of $J$.
Here each system has 4 electrons, while 
the size of the system is varied from 8 (Fig.\ref{224}), 
10 (Fig.\ref{226}) to 12 
(Fig.\ref{228}).
We can immediately notice three features \cite{phi05}: 
\begin{itemize}
\item The extended period, which is originally $N_a \Phi_0$ at $J=0$, 
is abruptly halved as soon as $J\neq 0$ is turned on. 
\item At a certain $J_C >2$, whose value depends on the band filling, 
a level anticrossing appears in the flow line. 
For 8- and 12-site systems, this makes 
the period further halved into a $1/4$ period compared to that for $J=0$. 
In the case of a $10(\neq 4, {\rm mod} 4)$ site system, the appearance of 
the 1/4 period is less perfect. 
\item Solvable points ($J=0, 2$) 
are distinguished by smooth, cosine-like flow lines. 
This sharply contrasts with the flow lines in the non-integrable case, 
where the lines wiggle due to ubiquitous level anticrossings.  
Despite these, the extended AB period does remain the 
same across the integrable-nonintegrable boundary, 
which is explicitly shown by the fact that the curvature of the 
anticrossing 
shows no singularities for all values of $J$ considered here. 
\end{itemize}

In the following sections, 
we explore why these features should arise. 
We divide the regions into exactly solvable points ($J=0$, $J=2$), 
and other regions (normal: $0<J<2$; 
possibly superconducting 
or phase-separated: $2<J$).

\subsection{$J=0$} 
Since the $t$-$J$ model with $J=0$ is equivalent 
to the $U=\infty$ Hubbard model, 
let us start with the 1D Hubbard model.
Consider an $N_a$-site ring (with $N_a$ even for simplicity) 
containing $N$ electrons. The Hamiltonian is given by 
\[
 H=-t\sum_{i \sigma} (c^{\dagger}_{i \sigma}c_{i+1 \sigma} + {\rm H.c.})
 +U\sum_{i}n_{i \uparrow}n_{i \downarrow}
\]
where $c_{i \sigma}$ ($c_{i \sigma}^\dagger$) 
annihilates (creates) an electron at the $i$-th 
site with spin $\sigma$, $n_{i\sigma}\equiv c_{i\sigma}^\dagger
c_{i \sigma}$, 
$U$ the on-site interaction, and $t$ is the transfer energy 
taken to be the unit of energy hereafter. 

We can analyze the 1D Hubbard model in terms of 
the exact Bethe-ansatz (BA) solution due to Lieb and Wu.\cite{lw} 
If we thread a magnetic flux $\Phi$ through the ring, 
the BA equations become\cite{ss} 
\begin{eqnarray}
 &&e^{ik_{j} N_{a}} = e^{i 2 \pi \Phi/\Phi_0} \prod^{M}_{\alpha=1} 
 e \left( \frac{4(\sin k_{j}-\lambda_{\alpha})}{U}\right) \, ,  \\
 &&\prod^{N}_{j=1} e \left( \frac{4(\lambda_{\alpha}-\sin k_{j})}{U}\right) 
 = \prod^{M}_{\beta=1} e\left( \frac{2(\lambda_{\alpha}
-\lambda_{\beta})}{U}\right) \, . 
 \label{1DHMBA}
\end{eqnarray}
Here $M (\leq N/2)$ is the number of down-spin electrons, 
$ e(x) \equiv (x+i)/(x-i)$, and 
${k_{j}}$ (${\lambda_{\alpha}}$) is 
the charge (spin) rapidities, which are roughly a quasimomenta for 
the charge (spin). 

When $U=\infty$,
the charge part of the BA equation (eq.(3.1)) becomes\cite{Ogata2} 
\begin{eqnarray*}
  k_j N_a&=&2\pi I_{j}+ 2 \pi \Phi/\Phi_0+P_H , \\
  P_H&=&\sum_{\alpha=1}^{M} 2 \tan^{-1}(2\Lambda_{\alpha}),
\end{eqnarray*}
with $\Lambda_{\alpha}=2 \lambda_{\alpha}/U$. 
This shows that in the strong-interaction limit the allowed 
values of $k_j$'s 
are quantized in units of $2\pi/N_a$. 
In other words, charge degrees of freedom act like spinless fermions. Hence 
the flow line has a periodicity of $N_{a} \Phi_0$. 

\subsection{$J=2$, the supersymmetric point}

$J=2$ is special in 
that the model is integrable with the Bethe-ansatz  solution, which 
we first briefly summarize.
Lai\cite{Lai} and later Schlottmann,\cite{Schlottmann} 
proved that solving the problem at $J=2$
reduces to solving a set of $(N+M)$-coupled algebraic 
equations for $N$ charge rapidities $v_{j}$ and $M$ spin rapidities
$\Lambda_{\alpha}$, i.e., 
\begin{eqnarray*}
\exp (i k_{j} N_{a})=\prod^{M}_{\beta=1} \frac
{v_{j}-\Lambda_{\beta}+i/2}{v_{j}-\Lambda_{\beta}-i/2}, \\
\prod^{N}_{j=1} \frac
{v_{j}-\Lambda_{\alpha}+i/2}{v_{j}-\Lambda_{\alpha}-i/2}
=-\prod^{M}_{\beta=1}\frac{\Lambda_{\beta}-\Lambda_{\alpha}+i}
{\Lambda_{\beta}-\Lambda_{\alpha}-i},
\end{eqnarray*}
with $v_{j}= \cot (k_{j}/2)/2$.
The ground state is given by a set of complex charge rapidities
(two-strings) and real spin rapidities,
where 
a spin rapidity lies at the center of each two string.

Sutherland\cite{Sutherland2} found an alternative 
way of solving the problem.  
There we treat the holes and down-spin electrons 
as dynamical objects in a background of up-spin electrons.
In this formalism, BA equations become
\begin{eqnarray*}
\exp (i k_{\alpha} N_{a})&=& -\prod^{M_2}_{j=1}
\frac{v_{\alpha}-w_{j}+i/2}{v_{\alpha}-w_{j}-i/2}
\prod^{M_1}_{\beta=1}
\frac{v_{\alpha}-v_{\beta}-i}{v_{\alpha}-v_{\beta}+i}, \\
1&=&\prod^{M_1}_{\beta=1}\frac{w_{j}-v_{\beta}-i/2}{w_{j}-v_{\beta}+i/2}.
\end{eqnarray*}
Here $v_{\alpha}=\tan(k_{\alpha}/2)/2$ 
with $\alpha=1,\cdots,M_{1}=N_{h}+M$ 
involves both spin and charge
degrees of freedom, while 
$w_{j}$ is the hole rapidity that represents charge degrees
of freedom 
with $j=1,\cdots, M_{2}=N_{h}$ with $N_{h}$ being the number of holes. 
A particle-hole transformation connects
Sutherland's representation to that of Lai or Schlottmann.
\cite{Bares2,comment}

Sutherland's formalism is more appealing for numerical calculations,
since 
(i) the ground state involves only real
roots, and 
(ii) we can directly investigate 
the behavior of the charge rapidities.
So we follow Sutherland.  
If we take the logarithm of Sutherland's BA equations we have
\begin{eqnarray*}
N_{a}\theta(2v_{\alpha})&=&2\pi J_{\alpha}+\sum_{\beta=1}^{M_1}
\theta(v_{\alpha}-v_{\beta})-\sum^{M_2}_{j=1}\theta(2(v_{\alpha}-w_{j})), \\
2 \pi I_{j}&=&
\sum^{M_{1}}_{\beta=1}\theta(2(w_{j}-v_{\beta})),
\end{eqnarray*}
with $\theta(x) \equiv 2 \tan^{-1}(x)$. 
$J_{\alpha}$ is an integer (half odd integer) when 
$M_{1}-M_{2}+1$ is even (odd), while $I_{j}$ is an integer 
(half odd integer) if $M_{1}$ is even (odd).  
Here we assume $I_{1}<I_{2}<\cdots <I_{M_2}$, $J_{1}<J_{2}<\cdots<J_{M_1}$.
For the ground state, we must choose $|J_\alpha|$'s and $|I_j|$'s 
as close to 0 as possible.

If we thread a magnetic flux $\Phi$, the left hand side of the
charge part of the 
BA equation becomes
\begin{eqnarray*}
2 \pi \tilde{I}_{j}\equiv 2 \pi I_{j}-2\pi \Phi/\Phi_0.
\end{eqnarray*}
In Fig.\ref{rapfig} we show how the rapidities evolve with 
$\Phi$ for the ground state. 
We can see that the rapidities change with  a characteristic 
manner with a periodicity of $N_a \Phi_0/2$.  
Conspicuously, 
the charge rapidities, $w_j$'s, respond to the magnetic flux 
{\it in pairs}. 
The pair diverges to $\pm \infty$ at 
some points, around which the pair sandwiches other rapidity 
$v_{\alpha}$.  
Since each pair of charge rapidities exchange their places as they 
cross from $-\infty$ to $+\infty$, 
the charge rapidities are shuffled like 
\[
\{ w_1,w_2,w_3,w_4,\cdots \} \rightarrow \{ w_2,w_1,w_4,w_3,\cdots \}
\]
as one extended period ($\Phi=0 \rightarrow \Phi=N_a\Phi_0/2$) 
is accomplished. 
The shuffle is similar to the case of the small $J$ limit, as we shall see  
in the next section.  We now look more closely at the 
BA equations.

\subsubsection{The charge part of the BA equation}
As $2 \pi \tilde{I}_{j}$ becomes smaller than $-M_{1}\pi$,
$w_{1}$ and $w_{2}$ vary from $-\infty$ to $\infty$.
At the same time, $v_{1}$ varies from $-\infty$ to $\infty$. 
In order to fix the range of $\tan^{-1}$,
we redefine the quantum numbers ${I_j}$ so that
$\{\tilde{I}_{1},\tilde{I}_{2},\cdots,\tilde{I}_{M_2}\}=
\{-M_{1}/2,-M_{1}/2+1 ,-M_{1}/2+2,\cdots,-M_{1}/2+M_2-1\}$
becomes $\{M_{1}/2-1,M_{1}/2,-M_{1}/2+1,\cdots,-M_{1}/2+M_2-2 \}$.
When this process is repeated, $\{\tilde{I}_{j}\}$ comes back to 
$\{I_{j}\}$ when $\Phi=N_a \Phi_0/2$.

\subsubsection{The spin part of the BA equation}
When $v_{1}$ varies from $-\infty$ to $\infty$, 
the quantum numbers will be redefined so that
$\{J_{1},J_{2},\cdots,J_{M_{1}}\}$ becomes
$\{J_{1}+M_1-1,J_{2}-1,J_{3}-1,\cdots ,J_{M_{1}}-1 \}$ 
for the present choice of the Riemann plane.
As a set, they do not change.

Accordingly the BA equations have a periodicity of
$N_a \Phi_0/2$, so that the rapidities return to the original values as a set
when $\Phi=N_a \Phi_0/2$.
Note that in this case, the spectral flow corresponds to a trajectory of the 
holon-antiholon excitations considered by Bares {\it et al}.\cite{BBO}

\subsection{$0<J<2$}

Remarkably, above halving of the extended AB period for a 
sizeable $J=2$ is seen to emerge 
already for an infinitesimal $J$ from the numerical result.  
In other words, the $t$-$J$ model with a finite $J$ and the Hubbard model have 
different extended AB periods.  
Here we show why an infinitesimal $J$ is enough to halve 
the extended AB period.  
For a small $J/t$ we can treat the 
$J$-term perturbatively. One can 
start from the $N$-particle state expanded as
\[
  |F\rangle = 
      \sum_{x_{1}}\cdots \sum_{x_{N}} \sum_{\sigma_{1}}\cdots \sum_{\sigma_{N}}
      f_{\sigma_{1}\cdots\sigma_{N}}(x_{1}\cdots x_{N})
      \prod c^{\dagger}_{x_{1}\sigma_{1}}\cdots
      c^{\dagger}_{x_{N}\sigma_{N}}|0\rangle \; . 
\]
For $J=0$, it is known\cite{Ogata2} that the coefficient $f$ factorizes into 
charge and spin parts as 
\[
  f_{\sigma_{1}\cdots\sigma_{N}}(x_{1}\cdots x_{N})
  =(-1)^Q{\rm det}[{\rm exp}(i k_{i} x_{Qj})] \Phi(y_{1}\cdots y_{M}) \; ,
\]
where $x_{Q1}<x_{Q2}<x_{Q3} \cdots <x_{QN}$ are the coordinates of 
all the electrons with $Q$ being a permutation and 
$k_{j}$ the momentum of a free spinless electron, while 
$y_{1}<y_{2} \cdots <y_{M}$ are the coordinates of down spins
in the spin configuration.  
The determinant depends only on the positions of particles 
$(x_{Q1},\cdots,x_{QN})$ and not on the positions of down spins 
$(y_{1},\cdots,y_{M})$. 
The spin part $\Phi(y_{1}\cdots y_{M})$ is identical to the wavefunction 
of the 1D Heisenberg ring of length $N$. 

When a small $J$ is turned on, characteristic change occurs 
at a special point in the flow, {\it i.e.,} highly degenerate points. 
Effect of $J$ first appears as a lifting of this degeneracy, 
which can be examined by the first-order degenerate perturbation, 
as is done\cite{Kusakabe} for the case of the 1D Hubbard model 
for $U\sim 0$. 
The effective Hamiltonian 
in the first-order perturbation becomes\cite{tJJ} 
\begin{eqnarray}
  H_{\rm eff}=-J \langle n|\sum_{i} n_i n_{i+1}|m \rangle_{SF} 
  \sum_{j}{\bf S}_{j} \cdot {\bf S}_{j+1},
\label{eq:Heff}
\end{eqnarray}
where $n_i=c^\dagger_i c_i$ is the number operator of spinless fermions, 
and $\langle n|\cdots|m \rangle _{SF}$ is the matrix element between 
wave functions of non-interacting spinless fermions having the 
same energy. 
The summation over $j$ is taken over the compressed spin chain 
(ignoring the hole sites). 
If we Fourier transform the first part of the effective Hamiltonian as 
\[
  -J\sum_{i}c_{i}^{\dagger}c_{i}c^{\dagger}_{i+1}c_{i+1}=
  -\frac{J}{N_a}\sum_{k,k',q} e^{{\rm i}q} 
  c^{\dagger}_{k-q}c_{k}c^{\dagger}_{k'+q}c_{k'},
\]
we can see that when $J$ is turned on, the degenerate states having
$\{ \cdots, k, \cdots, k', \cdots \}$ 
can mix to give an anti-crossing with the states having 
$\{ \cdots, k-q, \cdots, k'+q, \cdots \}$.  
For these states to have the same energy, 
we have $(k+k')/2\equiv \pm \pi/2({\rm mod} 2\pi)=\pi/2$ or $3\pi /2$.

We can compare in Figs.\ref{224} $\sim$ \ref{228} the numerical 
result for $J=0.1$ with that 
for $J=0$ obtained by the Bethe-Ansatz assuming $q=\pi$.
It can be seen that the spectral flow at $J=0.1$ 
delineates the lower envelope of the
$J=0$ flows, suggesting that we have indeed $q=\pi$.
For example, let us consider the case of 4 electrons in 12 sites
(Fig.\ref{228}).
When $J$ is turned on, the rapidities $\{k_j\}$ change as 
\[
\{ \frac{\pi}{12},\frac{3\pi}{12},\frac{5\pi}{12}, \frac{7\pi}{12} \} 
\rightarrow
\{ \frac{\pi}{12}, \frac{3\pi}{12},\frac{5\pi}{12}-\pi 
= \frac{-7\pi}{12},\frac{7\pi}{12}+\pi \equiv \frac{-5\pi}{12} \}
\]
at $\Phi=2.5\Phi_0$, and 
\[
\{ \frac{5\pi}{12},\frac{7\pi}{12},\frac{-3\pi}{12}, \frac{-\pi}{12} \} 
\rightarrow
\{\frac{5\pi}{12}-\pi = \frac{-7\pi}{12},\frac{7\pi}{12}+\pi 
\equiv \frac{-5\pi}{12},\frac{-3\pi}{12}, \frac{-\pi}{12} \} 
\]
at $\Phi=4.5\Phi_0$. 

We can indeed give a perturbational argument why 
the umklapp ($q=\pi$) is selected. 
For $J=0$, the state 
$\displaystyle
| \frac{\pi}{12},\frac{3\pi}{12},\frac{5\pi}{12}, 
\frac{7\pi}{12}\rangle \equiv |0\rangle$
is degenerate at $\Phi =2.5\Phi_0$ with the states
$\displaystyle
| \frac{\pi}{12},\frac{3\pi}{12},\frac{5\pi}{12}-q, 
\frac{7\pi}{12}+q \rangle \equiv |q\rangle$, where $q$ is 
one of vectors, 
$\displaystyle
\frac{3\pi}{6},\frac{4\pi}{6},\frac{5\pi}{6},\frac{6\pi}{6},
\frac{-4\pi}{6}$ or $\displaystyle \frac{-5\pi}{6}$.

The matrix for $\sum n_{i}n_{i+1}$ appearing in eq.(\ref{eq:Heff}) 
spanned by these states are
\[
\bordermatrix{
& q=0 & {6\pi}/{6} & {5\pi}/{6} & {4\pi}/{6}
& {6\pi}/{6}& {-4\pi}/{6}& {-5\pi}/{6}
\cr
q=0 & 1 & a^{-6} & a^{-5}& a^{-4} & a^{-3} & a^{4} & a^{5} \cr
{6\pi}/{6} & a^{6} & 1 & a & a^2 & a^3 & a^{10} & a^{11} \cr
{5\pi}/{6} & a^{5} & a^{-1} & 1 & a & a^2 & a^{9} & a^{10} \cr
{4\pi}/{6} & a^{4} & a^{-2} & a^{-1} & 1 & a & a^{8} & a^{9} \cr
{3\pi}/{6} & a^{3} & a^{-3} & a^{-2} & a^{-1} & 1 & a^{7} & a^{8} \cr
{-4\pi}/{6} & a^{-4} & a^{-10} & a^{-9} & a^{-8} & a^{-7} & 1 & a \cr
{-5\pi}/{6} & a^{-5} & a^{-11} & a^{-10} & a^{-9}& a^{-8}&a^{-1}& 1 \cr
}\]
where $a\equiv \exp (i\pi/6)$. 
This matrix has one nonzero eigenvalue, 
7, with the eigenvector $|m\rangle \equiv (1,a^6,a^5,a^4,a^3,a^{-4},a^{-5})$, 
while all the other eigenvalues are zero.
Hence, when a small $-J\sum_{i} n_{i}n_{i+1}$ 
is turned on, the degeneracy is partially
removed with a single level peeled off (Fig.\ref{sflow}). 
For the flows to be connected continuously, 
the state $|0\rangle$ 
must transform as
$|0 \rangle \rightarrow |m \rangle \rightarrow |q=\pi \rangle$, which 
exactly implies $q=\pi$. The above argument holds for any numbers of 
electrons and sites.
Interestingly, this behavior is similar to that for the attractive
Hubbard model, in which $q=\pi$ holds as well.

The motion or the `selection rule' for the rapidities 
can be seen more directly 
in Fig.\ref{chrap}, where 
two $k$'s pair off and respond to $\Phi$ hand in hand.  
We can interpret this as an effect of $J$ working as an attractive interaction.
As a consequence, a finite $J$ halves the period of the spectral flow. 
Namely, the $k$'s 
for the ground state 
change from $\{ k_1, k_2, k_3, k_4,\cdots,\}$ to
$\{ k_2, k_1, k_4, k_3,\cdots,\}$ for $\Phi = 0 \rightarrow N_a \Phi_0/2$.

This behavior is reminiscent 
of the charge rapidities in Fig.6, 
which can be defined when the system is integrable at 
$J=2$, the supersymmetric case.  
For $J=2$, 
Bares {\it et al.}\cite{BBO} have termed the phase as 
`weakly bound electron pairs' having antiparallel spins
because charge rapidities by
Lai's representation form two strings. 
In terms of the extended AB period, the numerical result shows that 
the period is the same over 
the whole region of $0<J<J_C$.  
In this sense the `weakly bound pairs' region extends over 
$0<J<J_C$.  

\subsection{$J>2$}
The present result, Figs.\ref{224} $\sim$ \ref{228},
exhibits a period halving 
at a critical $J_C(>2)$.  
In order to determine $J_C$, we have evaluated the energy gap $\Delta$
in the following way. 
First, we identify, with a linear extrapolation, 
the point $E_0$ at which levels would cross 
if the level repulsion were absent.  
The energy gap is then estimated as 
$\Delta=E_0-E_1$, 
where $E_1$ is the energy of the repelled level.
We can next plot $\Delta$ as a function of $J$ and determine
$J_C$ as $\Delta(J_C)=0$ by extrapolation.
The results are $J_C \simeq 2.30$ for 4 electrons in 8 sites,
$J_C \simeq 2.22$ for 4 electrons in 10 sites,
$J_C \simeq 2.14$ for 4 electrons in 12 sites.

According to the phase diagram obtained by Ogata {\it et al},\cite{Ogata}
the superconducting correlation function becomes dominant ($K_\rho>1$) 
in the region $J>J_c$, where $J_c>2$ 
is a function of the band filling. 
The phase boundary 
is identified as the trajectory of $K_\rho=1$ 
in terms of the exponent  $K_\rho$ that 
characterizes the TL liquid. 
The boundary between superconducting and phase separated regions
is more or less 
pararell to the trajectory of $K_\rho=1$.

Figure \ref{ogata} superposes the period halving points 
on the phase diagram (trajectory of $K_{\rho}=1$) the TL 
result\cite{Ogata}.
We can see the halving points fall on a line near the trajectory
of $K_\rho=1$, which is pararell to the boundary line
between superconducting and phase separated regions.

\section{$t$-$J$ ladder model}

We now turn to the $t$-$J$ ladder model 
(for a review of the ladder, see, e.g., Ref.37). 
The original motivation to introduce ladders came from an 
expectation that a gap in spin excitations may be formed 
in such systems, leading possibly superconductivity when carriers
are doped.  
The importance of the spin gap has in turn conceived for 
the high-$T_c$ cuprates.  Specifically, 
the $t$-$J$ ladder has been predicted to 
have superconductivity associated with a spin gap 
when there are even number of chains\cite{Dagotto92,Rice,Takashi}.  
The interests are heightened when a class of cuprates 
are found to possess the ladder structure 
in the Cu-O network.\cite{Mccarron,Siegrist,Hiroi} 
Very recently, a superconductivity has been detected in 
(Ca,Sr)$_{14}$Cu$_{24}$O$_{41}$ under high pressures.\cite{Uehara,Takahashi} 

This has kicked off numerical works with 
the exact diagonalization or the density-matrix renormalization group 
(DMRG) 
method.
\cite{Dagotto92,Tsunetsugu94,Tsunetsugu95,Poilblanc95,Hayward95,Troyer95} 
Since the system is quasi-1D, the Tomonaga-Luttinger (TL) 
liquid theory has also been applied.\cite{Hayward96,Sano96,Tsunetsugu95} 
Some results have predicted existence of a superconducting phase, 
in which a spin gap is also shown to survive the doping near 
the half filling.\cite{Tsunetsugu94} 
However, the ladder is not purely 1D, the validity of the
TL approach has to be justified.
Hence, another method detecting the pairing is desirable.

Thus it is most intriguing how the extended AB period test 
predict the transition for the $t$-$J$ ladder. 
Kusakabe and Aoki \cite{Kusakabe} obtained a preliminary result that
the extended AB period changes 
around the phase boundary predicted by Sano.\cite{Sano96} 
We discuss the extended AB period of this model
in more detail. 

The $t$-$J$ ladder is defined, in standard notations, by the Hamiltonian, 
\begin{eqnarray}
H&=&-t\sum_{i,\alpha,\sigma} P_G(c^\dagger_{i,\alpha,\sigma} 
c_{i+1,\alpha,\sigma}+ {\rm h.c.})P_G -t_{\perp}\sum_{i,\sigma} P_
G
(c^\dagger_{i,1,\sigma}c_{i,2,\sigma}+ {\rm h.c.})P_G \nonumber \\
&+&J\sum_{i,\alpha}({\bf S}_{i,\alpha}\cdot {\bf S}_{i+1,\alpha} 
- \frac{1}{4} n_{i+1,\alpha}n_{i,\alpha}) \nonumber \\
&+&J_{\perp}\sum_{i}({\bf S}_{i,1}\cdot {\bf S}_{i,2} 
- \frac{1}{4} n_{i,1}n_{i,2}) , 
\label{tjhami}
\end{eqnarray}
where $\alpha$ labels two legs of the ladder, while 
$i$ labeling the rung runs from 1 to $N_a$. 
Doubly occupied sites are totally excluded 
by the Gutzwiller projection $P_G$  as in the single $t$-$J$ model. 

Here we take the same intra- and inter-chain electron 
transfer, $t=t_{\perp}$, and 
the same intra- and inter-chain 
superexchange interaction, $J=J_{\perp}$, for simplicity. 
The AB flux threaded to the ladder wound
along the length introduces a Peierls phase as 
\begin{equation}
c^\dagger_{i,\alpha,\sigma} c_{i+1,\alpha,\sigma} \rightarrow
{\rm exp}\left( i\frac{2\pi\Phi}{N_a\Phi_0}\right) 
c^\dagger_{i,\alpha,\sigma} 
c_{i+1,\alpha,\sigma}.
\end{equation}
We have numerically 
obtained the extended spectral flow for finite $t$-$J$ ladders. 
Results depends on the number of electrons. 

\subsection{Two-Electron Systems}

Result for two electrons in an $N_a\times 2 = 8\times 2$-site system 
is shown in Fig.\ref{tjlad2}
We look at the $J$-dependence of the flow of the ground state, 
starting from the $t$ model (the system with no superexchange interactions, 
$J=0$, but the double occupancies still excluded).  
For $J=0$ the flow line has a periodicity of $N_a \Phi_0$, 
which is the period expected for individually moving particles. 
Remarkably, a small $J(=0.4t)$ is enough to halve the periodicity. 
This behavior, which is common to two-electron systems on 
$2n \times 2$-site ladder ($n$: an integer), 
is caused again by a pair of level anticrossings in the flow. 
The change in the period from $N_a \Phi_0$ to $N_a \Phi_0/2$ 
is similar to the result for the single-chain $t$-$J$ model (Section 3), 
in which the period is also halved for an infinitesimal $J$.
For general numbers of electrons in a ladder, however, 
the period $N_a \Phi_0$ is not always observed 
even for $J=0$. 
For some numbers of electrons, 
the halved period, $N_a \Phi_0/2$, is often detected (see below) 
for the $t$ ladder, 
although $t$ model has 
no explicit source of attractions between electrons. 

\subsection{Four-Electron Systems}

We move on to the spectral flows for four electrons 
for an $8\times 2$-site system (1/8-filling) 
in Fig.\ref{tjlad4-1} 
for various values of $J$. 
We can see that 
there exist 
a pair of level anti-crossings 
at $J_{C} (\simeq 2.6t$), 
which reduces the extended period into $N_a \Phi_0/4$.

For a $6\times 2$-site system 
(1/6-filling) in Fig.\ref{tjlad4-2}, 
a similar appearance of a pair of level anticrossings occurs 
at $J_{C}\simeq 2.04t$,\cite{Kusakabe} 
although the period halving is less ideal.  
This can be regarded as a pseudo 1/4-periodicity, in which 
the exact 1/4 period is inhibited due to the 
incommensurability of $N_a$ with 4. 

In Fig.\ref{tjladphase} we compare  the $J_C$ obtained from 
the extended period halving with 
the phase diagram obtained from the TL analysis by 
Hayward {\it et al.}\cite{Hayward96} and independently by Sano
\cite{Sano96}. 
The value of $J_C$ falls on a line near the trajectory of
$K_\rho=1$ and nearly pararell to the
boundary between superconducting and
phase-separation regions.

Here we should comment the following.  
The period is already halved at $J=0$ (the $t$ ladder) 
at least for the $6\times 2$ system.\cite{comm82} 
This should be related 
to the Hubbard ladder (the Hubbard model on a double chain), 
since the $t$ model is identical to the Hubbard model with $U=\infty$. 
Thus the Hubbard ladder should have the halved period as well 
for $U=\infty$, while 
for $U=0$ the extended period is $N_a \Phi_0$. 
This implies that a critical point {\it has to} exist at which 
the extended period changes in the Hubbard ladder. 
A preliminary calculation for a $6\times 2$-site Hubbard ladder 
with 4 electrons indeed indicates a halving of the extended period at 
a finite $U_c$. 
If this is true, we have another example in which the property of 
the ladder is distinct from the single chain, 
since the period is always $N_a \Phi_0$ in the single-chain 
Hubbard model with $U>0$. 

\subsection{Quarter filling}

At the quarter filling ($N_a$ electrons on a $N_a\times2$ ladder), 
a peculiar behavior appears, {\it i.e.}, the 
period becomes the smallest possible 
$\Phi_0$ irrespective of $J$ and the systems size 
($4\times 2$, $6\times 2$ and $8\times 2$ sites). 
As an example we show the ground-state spectral flow for a $6\times 2$-site 
system with 6 electrons in Fig.\ref{tjlad6}. 
The period $\Phi_0$ is what is expected for the charge-gapped system 
(e.g., a Mott insulator). 
In this sense the 
$\Phi_0$ period may come from the commensulability of the filling, 
although we could not find any characteristic structure in 
the CDW correlation $\langle n_{i,\alpha} n_{j,\beta} \rangle$, 
which is almost flat in these small systems. 
The period remains the same when we modify the interaction parameters 
in the region $t < t_\perp$, $J < J_\perp$. 
On the other hand, if we approach to the independent chains in the region 
$t > t_\perp$, $J > J_\perp$, 
a branch having a longer period 
comes downwards to cross the lowest $\Phi_0$-period line. 
This is natural because in the limit of $t \gg t_\perp$, $J \gg J_\perp$, 
the extended period for the ground state should be 
that for the 1D ring with $N/2$ electrons.

\section{$t$-$J$-$J'$ model}

Ogata {\it et al} have studied 1D $t$-$J$-$J'$ model, \cite{tJJ} where
they added a frustrating next nearest-neighbor exchange interaction
$J'$ to the $t$-$J$ model.
Like ladder systems, this has been 
motivated from an expectation that we can tune the 
spin gap by introducing $J'$.  
$J'$ is in fact shown to give
rise to a spin gap at half-filling,
which survives for small doping and arbitrary values of $J$. 
In this section, we apply our method to $t$-$J$-$J'$ model 
to discuss whether the extended AB period can detect 
a symptom of changes in characters of 
quantum state.

In particular,
the halving point, $J_C$, of the extended AB period from 
$N_a \Phi_0/2$ to $N_a \Phi_0/4$ is of interest. 
(As in 1D $t$-$J$ model, the extended AB period halves to $N_a \Phi_0/2$
right after the $J$ and $J'$ set in.)
We fix $J'/J=0.5$, because Ogata {\it et al}
\cite{tJJ} determined the phase diagram for this case.  
The present result shows that 
$J_C=3.4t$ for 4 electrons in 8 sites, 
$J_C=2.8t$ for 4 electrons in 10 sites, and 
$J_C=2.6t$ for 4 electrons in 12 sites.
In Fig.\ref{tjj}, we superpose the present result on the phase diagram
due to Ogata {\it et al} \cite{tJJ}.
We can see that $J_C$ determined by the extended AB effect 
falls near the neighborhood 
of the boundary of phase separation obtained 
by the exact diagonalization method.  

\section{Extended Hubbard model}
So far we have studied models for which the phase diagram
contains spin-gapped regions. However, we have not studied
the behavior of the AB period in such regions so far. 
It is interesting to investigate
how the extended AB period behaves in
the region where spin gap opens, 
since the opening of a spin gap
is regarded as a tendency toward attraction
between electrons.  
In the extended Hubbard model, where 
we consider the off-site (nearest-neighbor) 
interaction $V$ in addition to the 
on-site Hubbard $U$, 
the phase diagram 
has both spin-gapped and spin-gapless regions bisected 
by the so-called Luther-Emery line.  
Thus the model is ideal for studying 
how the existence or otherwise of the spin gap 
affects the extended AB effect.  

The extended Hubbard model is defined by the Hamiltonian,
\[
H=-t\sum_{<i,j>\sigma} \left( c^{\dagger}_{i\sigma} c_{j\sigma}
+ {\rm H.c.}\right)
+U\sum_{i}n_{i\uparrow}n_{i\downarrow}
+V\sum_{i\sigma}n_{i\sigma}n_{i+1\sigma}.
\]
According to the weak-coupling theory\cite{}, the superconducting correlation
dominates in the region
\[
U+2[2-\cos(2k_F)]V <0
\]
and the spin gap opens in the region 
\[
U+2V\cos(2k_F)<0
\]

For the strong-coupling region, the phase diagram has to be 
obtained numerically.  
Making use of the method same as the 1D $t$-$J$ study by Ogata {\it et al}
\cite{Ogata}, Sano and {\=O}no \cite{Sanoeh},
and independently three of the present authors \cite{kurokieh},
have numerically determined the phase diagram. 

Here we focus on the region with $V<0$ 
for which the pairing is expected.  
When $V$ set in the extended AB period halves to $N_a \Phi_0/2$ 
regardless of the value of $U$.
We have determined the point where the extended AB period is halved 
from $N_a \Phi_0/2$ to $N_a \Phi_0/4$ for 4 electrons in 8 sites,
4 electrons in 10 sites, and 4 electrons in 12 sites.  
In Fig.\ref{4ele}, we superpose our results on the phase diagram
due to Sano and {\=O}no \cite{Sanoeh}.
The halving point of the extended AB period is seen to 
run roughly parallel to the normal-superconductivity 
and superconductivity-phase separation
boundaries, and falls on the superconductivity region.\cite{vposi}

\section{Concluding Remarks}

In four types of strongly correlated electron systems in 1D,
namely
\begin{enumerate}
\item
$t$-$J$ model,
\item
$t$-$J$ ladder,
\item
$t$-$J$-$J'$ model,
\item
extended Hubbard model,
\end{enumerate}
we have obtained the extended AB period both from 
the Bethe-ansatz analysis where possible, 
and from numerical calculations.   
While the extended AB period is the system length times the flux quantum 
for noninteracting systems, 
we have found the existence of the boundary across which 
the period is halved or another boundary into an even shorter 
period on the phase diagram for these models.   
If we compare this result with the phase diagram predicted from 
the Tomonaga-Luttinger theory, 
the halved period (or shorter periods) has a one-to-one 
correspondence to the 
pairing (phase separation or metal-insulator 
transition) in these models.  
Namely, when the halved (quarter) AB period is detected, 
the corresponding TL phase diagram 
has a paired (phase-separated) region.  
Such an applicability of the extended AB test extends to 
the cases where the conventional anomalous flux
quantization gives ambiguous results as is the case with 
the Hubbard or $t-J$ models.

Surprising aspects of these results are:
\begin{enumerate}
\item
The extended AB flows involve high-energy states,
while the TL theory is only intended for the low-energy physics.
\item
The extended AB period does not change across
the integrable-nonintegrable boundary despite
the totally different level statistics.
\end{enumerate}

An important future problem is how the extended AB analysis 
can be coupled with {\it Berry's geometrical phase}.  
Korepin and Wu\cite{Korepin} have in fact calculated 
for the XXZ Heisenberg magnet Berry's phase from the Bethe-ansatz 
solution based on the periodicity ($\Phi=2\Phi_0$) found by 
Sutherland and Shastry.  
Berry's phase, $\gamma$, is expressed in a standard way as 
\begin{equation}
\gamma = {\rm Re} \left[ i \int^{\Phi_0}_{-\Phi_0} d\Phi 
\frac{\langle \psi(\Phi)|\frac{\partial}{\partial \Phi}|\psi(\Phi) \rangle}
{\langle \psi(\Phi)|\psi(\Phi) \rangle} \right].
\end{equation}
We can in principle calculate Berry's phase for electron systems 
such as the Hubbard model similarly.  
The geometrical phase will help identifying the nature of 
the many-body states in a gauge field theoretic manner.

\section{Acknowledgments}

We wish to thank Prof. V.E. Korepin for drawing our attention to
Ref.52 and Prof. N. Kawakami, Dr. X. Zotos and
Dr. H. Asakawa for valuable discussions.  

Numerical calculations were performed on FACOM VPP500 in
Supercomputer Center, Institute for Solid State Physics,
University of Tokyo. This work was supported in part by
a Grant-in-Aid for
Scientific Research on Priority Areas from
the Ministry of Education, Science and
Culture, Japan.

\begin{figure}
  \caption{An Aharonov-Bohm flux pierced in an opening of a ring.}
  \label{torus}
\end{figure}

\begin{figure}
  \caption{The extended AB flow for a 10-site Hubbard ring
    with 6 electrons$^{12}$. 
    The flow lines represent a repulsive interaction
    ($U>0$), an attractive interaction ($U<0$),
    and the noninteracting case.
    }
  \label{kusakabe}
\end{figure}

\begin{figure}
    \caption{Numerical result for the spectral flow for a 1D $t$-$J$ 
      ring with 4 electrons in 8 sites.}
    \label{224}
\end{figure}

\begin{figure}
    \caption{Numerical result for the spectral flow for a 1D $t$-$J$ 
      ring with 4 electrons in 10 sites.}
    \label{226}
\end{figure}

\begin{figure}
    \caption{Numerical result for the spectral flow for a 1D $t$-$J$ 
      ring with 4 electrons in 12 sites.}
    \label{228}
\end{figure}

\begin{figure}
  \caption{The rapidities for the hole charge ($w_j$: thin lines) and for
spin-charge degrees of freedom ($v_\alpha$: bold lines) against the
flux for the ground state for 
a 1D $t$-$J$ ring with 4 electrons in 10 sites. 
Tiny gaps in the non-diverging 
branches correspond to the points 
where a pair of branches diverge to $\pm \infty$.
}
  \label{rapfig}
\end{figure}

\begin{figure}
  \caption{The change of the spectral flow when small $J$ is 
    turned on.}
  \label{sflow}
\end{figure}

\begin{figure}
  \caption{The motion of charge rapidities for a finite $J$ in
1D $t$-$J$ model.}
  \label{chrap}
\end{figure}

\begin{figure}
  \caption{The present result for the halving points (open circles)
of the extended AB period in 1D $t$-$J$ model is superposed on the
phase diagram obtained by Ogata {\it et al.}$^{18}$ }
  \label{ogata}
\end{figure}

\begin{figure}
    \caption{The extended AB flow for an 8 $\times$ 2-site $t$-$J$ ladder
with two electrons for $J=0$, $J=0.4t$ 
and $J=0.8t$, respectively.}
    \label{tjlad2}
\end{figure}

\begin{figure}
    \caption{The extended AB flow for an 8 $\times$ 2-site $t$-$J$ ladder
with four electrons for $J/t=0,2.0,2.6,3.0,3.4$ from top to bottom.}
    \label{tjlad4-1}
\end{figure}

\begin{figure}
    \caption{The extended AB flow for an 6 $\times$ 2-site $t$-$J$ ladder
with four electrons for $J=0,2.0,2.2,2.6,\cdots,3.8$ from top to bottom}
    \label{tjlad4-2}
\end{figure}

\begin{figure}
  \caption
{
Halving points, $J_C$, of the extended AB period 
(big crosses) in the $t$-$J$ ladder 
are compared with the trajectory of ${K}_\rho=1$ 
(the normal-superconducting boundary) 
obtained by 
the exact diagonalization 
(solid squares by Hayward and Poilblanc$^{25}$,
solid triangles by Sano$^{26}$).
Phase separation boundary (${K}_\rho=\infty$) 
are also displayed by open squares or open triangles.
}
  \label{tjladphase}
\end{figure}

\begin{figure}
    \caption{
    The extended AB flow for an 6 $\times$ 2-sites $t$-$J$ ladder
with six electrons (the quarter-filling) for $J=0$ 
and $J=t$. 
The period is $\Phi_0$ irrespective of $J$. 
    }
    \label{tjlad6}
\end{figure}

\begin{figure}
  \caption{The present result for the halving points (filled circles)
of the extended AB period in 1D $t$-$J$-$J'$ model is superposed on the
phase diagram obtained by Ogata {\it et al.}$^{22}$ }
  \label{tjj}
\end{figure}

\begin{figure}
  \caption{The halving point 
of the extended AB period 
in the extended Hubbard model 
for four electrons in an 8-site system 
(band filling $n=0.5$; open circles), 
10-site system ($n=0.4$; open triangle), and 
12-site system ($n=0.333$; open square) is 
superposed on the 
phase diagram obtained by Sano {\it et al.}$^{23}$ 
and Kuroki {\it et al.}$^{24}$ for $n=0.5$.  
For $n=0.5$, The line $U=0$ is the boundary between the spin-gapped and
spin-gapless regions.
}
  \label{4ele}
\end{figure}


\begin{references}
\bibitem{comm82} For the $8\times 2$ system we cannot obtain the whole 
spectral flow at $J=0$ because of a numerical difficulty.
\bibitem{vposi} According to the calculation by Sano and {\=O}no,
there is a superconducting and phase-separation region in 
the region $V>0$ as well.
Our preliminary calculation indicates another transition point from
$N_a\Phi_0/2$ to $N_a\Phi_0/4$ in $V>0$ 
in qualitative agreement with this result.
\end{references}
\end{document}